\documentclass[12pt]{article}
\RequirePackage[OT1]{fontenc}
\usepackage{amsthm,amsmath,natbib}
\RequirePackage[colorlinks,citecolor=blue,urlcolor=blue]{hyperref}
\usepackage{amssymb}
\usepackage{graphicx,overpic,amsmath}
\usepackage{bm}
\usepackage{upgreek}
\usepackage{url}
\usepackage{color}
\usepackage[overload]{textcase}
\usepackage[ruled]{algorithm2e}
\usepackage{makecell}
\usepackage{natbib}
\usepackage{multirow}

\usepackage{subcaption}
\usepackage[shortlabels]{enumitem}
\usepackage[flushleft]{threeparttable}

\title{Nested Model Averaging on Solution Path for High-dimensional Linear Regression\footnote{ This work is partially supported by NSF CAREER Grant DMS-2013789. Feng and Liu contribute equally to this work}}

\author{Yang Feng$^{1}$ and 
	Qingfeng Liu$^{2}$\\ \\
	$^1$ New York University \\
	$^2$ Otaru University of Commerce
}

\date{}

\begin{document}
\baselineskip 16pt
\maketitle

\begin{abstract}
  We study the nested model averaging method on the solution path for a high-dimensional linear regression problem. In particular, we propose to combine model averaging with regularized estimators (e.g., lasso and SLOPE) on the solution path for high-dimensional linear regression.   In simulation studies, we first conduct a systematic investigation on the impact of predictor ordering on the behavior of nested model averaging, then show that nested model averaging with lasso and SLOPE compares favorably with other competing methods, including the infeasible lasso and SLOPE with the tuning parameter optimally selected. A real data analysis on predicting the per capita violent crime in the United States shows outstanding performance of the nested model averaging with lasso. 

\end{abstract}


\section{Introduction}
In the past two decades, a large amount of high-dimensional data sets are generated as a result of technological advancements in many fields.  Such data are characterized by a large number of total predictors compared with the available sample size.  For an overview of the many challenges and development associated with high-dimensional statistical modeling, we refer the readers to \cite{fan2010selective}  and \cite{buhlmann2011statistics}.

A crucial goal in high-dimensional data analysis is to strike a good balance between the goodness-of-fit and the complexity of the model, since both predictability and model interpretability are important to practitioners in many scientific fields. One popular avenue to achieve this balance is the imposition of regularization, which leads to simultaneous variable selection and parameter estimation in one single step. Some 
prominent examples include lasso \citep{tibshirani1996regression}, SCAD \citep{fan2001variable}, adaptive lasso \citep{zou2006adaptive},  MCP \citep{zhang2010nearly} among others. Recently, motivated by controlling the false positive rate, \cite{bogdan2015slope} proposed the SLOPE, where the $L_1$-sorted norm is used in the penalty form.

 For those regularized estimation methods, there has been abundant research \citep{zhao2006model, wainwright2009sharp, zhang2010nearly}  on their theoretical properties in various aspects. Those attractive properties usually require us to properly specify the penalty parameter, the determination of which generally depends on some unknown quantities. As a result, a data-driven choice of the penalty parameter under high-dimensional settings has been an important research question.  It is widely acknowledged that the traditional cross-validation and classic information criteria AIC and BIC may not perform well in high-dimensional scenarios. Some new tuning parameter selection methods tailed for high-dimensional settings were developed \citep{chen2008extended, fan2013tuning, feng2018restricted}.  
 
 An alternative approach to tuning parameter selection, or more generally model selection, is through model averaging. For linear models, \cite{hansen2007least} proposed Mallows model averaging (MMA) for nested models and showed it is asymptotically optimal in the sense of achieving the lowest possible squared error in a class of discrete model average estimators. \cite{wan2010least} extended MMA to handle non-nested models and showed the optimality of MMA hold for continuous model weights. \cite{Liu2013Heteroscedasticity} and \cite{liu2016generalized} extended MMA to linear regression models with heteroscedastic errors. \cite{zhang2016optimal} studied the optimal model averaging for generalized linear models and generalized linear mixed-effect models. For high-dimensional data analysis, \cite{FENGetal_sparsity_2020} developed a new algorithm to admit a large number of candidate models. The idea of model averaging has extended to various areas such as instrumental variable estimation \citep{Kuersteiner2010Constructing}, factor-augmented regression \citep{cheng2015forecasting}, quantile regression \citep{lu2015jackknife}, semiparametric ultra high-dimensional models, GARCH-type models \citep{LiuYaoZhao2020} and so on. See \cite{moral2015model} and \cite{steel2017model} for overviews of model averaging in economics and a book treatment on comparing model averaging and model selection in \cite{claeskens2008model}.

The main contribution of this work is two-fold. First, we propose to couple the nested model averaging method with regularization methods including lasso and SLOPE, and demonstrate that this coupling is very effective that leads to a smaller empirical risk than competing methods, including lasso and SLOPE with their tuning parameters optimally selected with the knowledge of true regression coefficients.  Second, we investigate the impact of predictor ordering on the behavior of nested model averaging and show the correct ordering help to reduce the in-sample loss substantially.

The rest of the paper is organized as follows. In Section \ref{sec::nma}, we introduce the nested model averaging framework and described the lasso-ma and slope-ma as two illustrating examples. Section \ref{sec::simulation} demonstrates the impact of variable ordering for model averaging estimates, and compared the model averaging models with competing model selection methods. In Section \ref{sec:real}, we compare the nested model averaging methods with the corresponding model selection methods on a real data for predicting per-capita violent crimes. We conclude the paper with a short discussion in Section \ref{sec:discussion}.

\section{Nested Model Averaging\label{sec::nma}}
Suppose we observe $n$ i.i.d. pairs $\{(X_i, Y_i)\}_{i=1}^n$ from $(x,y)$ where 
\begin{align}\label{eq:lm}
    y=\mu +\epsilon = x'\beta+\epsilon,
\end{align}
in which $\beta$ is the true regression coefficient, $x$ is the $p$-dimensional feature vector, and $\epsilon$ is the random error with $E\epsilon = 0$ and $E\epsilon^2=\sigma^2$. Let $X=[X_1', X_2',\cdots,X_n']'$ be  the $n\times p$ design matrix  and $Y=[Y_1,Y_2,\cdots, Y_n]'$ the $n\times 1$ response vector.  For a given coefficient vector estimate $\hat\beta$, we define the \emph{loss} function and the \emph{risk} function as $L_n(\hat\beta) = \|X(\hat\beta-\beta)\|_2^2$ and $R_n(\hat\beta) = EL_n(\hat\beta)$. Our goal is to find $\hat\beta$ such that $L_n(\hat\beta)$ and $R_n(\hat\beta)$ are as small as possible. 

Suppose we are given a total of $K$ candidate regression models as $\mathcal{M}=\{M_1,\cdots,M_K\}$, where the corresponding estimate for $\mu$ is $\hat\mu_k$ for $M_k$, for $k=1,\cdots, K$.  The idea of frequentist model averaging \citep{hansen2007least, wan2010least, zhang2016optimal} is to consider the following weighted average of those $K$ estimate: $\hat \mu(x, w) = \sum_{k=1}^K w_k \hat\mu_k$, where $w=(w_1,\cdots, w_K)^T$ is a weight vector in the unit simplex in $R^K$:
 \begin{align}
     \mathcal{H}_{K}=\left\{ \left.w\right|w\in\left[0,1\right]^{K}, \sum_{k=1}^{K}w_{k}=1\right\}. 
 \end{align}
 Here, we consider the $k$-th candidate model $M_k$ to be the linear model in \eqref{eq:lm} with regressors in $S_k\subset \{x_1,x_2,\cdots,x_p\}$, for $k=1,\cdots,K$. In addition, we define $s_k=\mbox{card}(S_k)$ as the number of predictors in $S_k$. We focus on the situation with nest candidate models, where $S_{k}\subset S_{k'}$ for $k<k'$.  For those $K$ candidate models, \cite{hansen2007least} and \cite{wan2010least} introduced the so called Mallows' Model Averaging (MMA) to choose $w\in \mathcal{H}_{K}$ that minimizes
 \begin{align}\label{eq:mma}
 C_n(w) = \sum_{i=1}^n (Y_i - \hat\mu(x,w))^2 + 2\hat\sigma^2s(w),
 \end{align}
 where $\hat\sigma^2$ is an estimate of $\sigma^2$ and $s(w) = \sum_{k=1}^K w_ks_k$ is the effective number of parameters for the model averaging estimate with weight vector $w$. \cite{wan2010least} showed such an estimate is asymptotically optimal in the sense of achieving the lowest possible loss among all model average estimators. 
 
 For a low-dimensional problem, one could consider all potential subsets to form candidate models and use the corresponding ordinary least square estimates. However, this strategy quickly becomes prohibitive when the dimension is moderately large (e.g., $p>50$) since the total number of models grows exponentially with $p$. As a result, researchers have advocated the use of MMA when the candidate models are nested \citep{hansen2007least, wan2010least}. These nested models correspond to a particular ordering of the predictors. Next, we describe two specific nested model averaging methods on solution paths generated from lasso (Section \ref{subsec:lasso-ma}) and SLOPE (Section \ref{subsec:slope-ma}), respectively.
 
%
%

\subsection{Lasso: Cross-validation and Model Averaging\label{subsec:lasso-ma}}
\cite{tibshirani1996regression} proposed the lasso procedure to simultaneously conduct variable selection and parameter estimation. The lasso estimator is given by
\begin{align}\label{eq::lasso}
\hat\beta_{lasso}(\lambda) = \arg\min_{\beta}\left\{\frac{1}{2}\|Y-X\beta\|_2^2+\lambda\|\beta\|_1\right\},    
\end{align}
where $\lambda$ is the regularization parameter. Under certain regularity conditions, lasso is able to achieve model selection consistency \citep{zhao2006model, wainwright2009sharp} when $\lambda$ is chosen properly.  In practice, one popular algorithm to calculate the lasso solution path is the \texttt{glmnet} algorithm \citep{friedman2009glmnet}, where $\lambda$ decreases from $\lambda_1=\lambda_{\max}$ to $\lambda_K=\eta\lambda_{\max}$, where $\lambda_{\max}$ is the maximum $\lambda$ that leads to a non-zero solution and $\eta$ is a small constant.\footnote{A typical choice when $p>n$ is $\eta=0.01$ and $K=100$ with the $\lambda$ sequence equal spaced in logarithmic scale.}  To achieve a small prediction error, one popular choice of the $\lambda$ is done through cross-validation. To remove potential bias, we follow \cite{belloni2013least} to conduct an ordinary least square estimate on the the active predictors corresponding to the lasso estimate with  $\lambda$  chosen by cross-validation. In particular, we consider the lasso-ols estimate as defined in 
\begin{align}
    \hat \mu_{lasso-ols} = X\hat\beta_{lasso-ols},
\end{align}
where $\hat\beta_{lasso-ols}= [X_{S(\lambda_{cv})}'X_{S(\lambda_{cv})}]^{-1}X_{S(\lambda_{cv})}'Y$, where $X_S$ represent the design matrix $X$ with column $S$, and $S(\lambda_{cv})$ is the support corresponding to the lasso estimate with $\lambda$ chosen by 10-fold cross-validation. It is shown in \cite{belloni2013least} that lasso-ols possesses better asymptotic property than the original lasso estimate. 

Although the strategy of choosing $\lambda$ by cross-validation works well for prediction under a low-dimensional setting,  a data-driven choice for $\lambda$ under high-dimensional settings remains an important open research question, and the optimal choice usually depends on the specific setup and the research question.  Naturally, we could perform model averaging on all available solutions on the solution path. 

To fix idea, for the sequence of lasso solutions indexed by $\lambda$, we apply model averaging method on those solutions using the coordinate-wise descent algorithm developed in \cite{FENGetal_sparsity_2020}. Note that the support of the solutions $\hat\beta_{lasso}(\lambda_k)$ for $k=1,\cdots,K$ are nested, which is in line with the nested model averaging framework. However, different from the original MMA \citep{hansen2007least}, we use the lasso solutions directly without further running an ordinary least square on the corresponding support. We observe this practice tends to provide more stable solutions than the lasso-ols based model averaging, especially when the dimension of the problem is relatively large. Now, we define the lasso-ma solution as 
\begin{align}\label{eq::lasso-ma}
\hat \mu_{lasso-ma} = X\left[\sum_{k=1}^K w^{lasso}_k \hat\beta_{lasso}(\lambda_k)\right],
\end{align}
where the weight vector $w^{lasso}=(w^{lasso}_1,\cdots,w^{lasso}_K)^T\in \mathcal{H}_{K}$ minimizes
\begin{align}
   \|Y - \hat \mu_{lasso-ma}\|_2^2 + 2\hat\sigma_{lasso}^2 \sum_{k=1}^K w_ks_k,
\end{align}
where $s_k=\|\hat\beta_{lasso}(\lambda_k)\|_0$ is the number of active predictors corresponding to solution $\beta_{lasso}(\lambda_k)$ and $\hat\sigma_{lasso}^2$ is the mean squared residuals corresponding to $\hat\beta_{lasso}(\lambda_{cv})$. Note that when lasso possesses the model selection consistency for certain $\lambda$, the solution path would correspond to the ``correct" ordering introduced in Section \ref{subsec::impact-ordering}, which would in turn lead to a small empirical risk. 

For comparison purposes, we would like to consider an ``infeasible" estimate on the lasso solution path, which gives us the smallest in-sample loss. More specifically, we define the lasso-optimal solution as 
\begin{align}\label{eq::lasso-optimal}
    \hat\mu_{lasso-optimal} = X\hat\beta^{lasso}_{k^*},
\end{align}
where
\begin{align}
    k^* = \arg\min_k \| X (\hat\beta^{lasso}_k - \beta)\|_2^2.
\end{align}
This ``infeasible" solution would give us the ``best" possible lasso estimate on the solution path if we know the true coefficient vector $\beta$. The motivation of considering this ``infeasible" estimate is that it shows the full potential of lasso estimate without looking at the performance corresponding to different ways to select the tuning parameter $\lambda$. In the simulation section, we will compare lasso-ols,  lasso-ma, and lasso-optimal in terms of in-sample loss. 

\subsection{SLOPE: False Discovery Rate Control and Model Averaging\label{subsec:slope-ma}}
When the model selection consistency conditions (e.g., the irrepresentable condition) for lasso hold, we expect the lasso-ma estimate to work well. However, it is observed in the literature that such conditions are very difficult to be satisfied in more practical settings \citep{su2017false}. In particular, \cite{su2017false} showed that on the lasso solution path, the false discoveries (i.e., the noise variables) usually appear before some important variables, which would, in turn, cause the lasso-ma to have a sizeable empirical risk. As a result, we would like to investigate some alternative regularization methods. 

 \cite{bogdan2015slope} proposed SLOPE (Sorted L-One Penalized Estimation), a regularization method to solve the high-dimensional linear regression problem, which aims to control the false discovery rate (FDR) at certain level. In particular, for any given FDR threshold $q\in(0,1)$, SLOPE is defined as 
\begin{align}\label{eq::slope}
\hat\beta_{SLOPE}(q) = \arg\min_{\beta} \left\{\frac{1}{2}\|Y-X\beta\|_2^2 + \sum_{j=1}^p \tau_j|\beta_{(j)}|\right\},
\end{align}
where $\beta_{(j)}$ is the $j$-th order statistic of $\beta=(\beta_1,\cdots,\beta_p)'$ and the penalty $\tau_j=z(1-j\cdot q/(2p))$, in which $z(\cdot)$ is the quantile of a standard normal distribution. This particular choice of weights enables us to control the FDR at given threshold $q$ \citep{bogdan2015slope}. \cite{su2016slope} showed that SLOPE is adaptive to unknown sparsity and is minimax optimal for a certain class of parameter space. Admittedly, one still need to choose the threshold $q$ to get a final estimate. Here, we propose to apply the model average method on the SLOPE solutions for a sequence of increasing $q$ values. In the numerical results, we consider the solution path $\{\hat\beta_{SLOPE}(q_k), k=1,\cdots, 9\}$, where $q_k = 10^{k-11}$. Here, we have in total $K=9$ SLOPE solutions.

Following a similar strategy as lasso-ols, we define SLOPE-ols as 
\begin{align}
    \hat\mu_{SLOPE-ols} = X\hat\beta_{SLOPE-ols}(q_{cv}),
\end{align}
where $\hat\beta_{SLOPE-ols}(q_{cv})= [X_{S(q_{cv})}'X_{S(q_{cv})}]^{-1}X_{S(q_{cv})}'Y$, where $S(q_{cv})$ is the support corresponding to the SLOPE estimate with $q$ chosen by 10-fold cross-validation.

We would like to introduce the model averaging estimate for the SLOPE solutions. In particular, we define the SLOPE-ma solution as 
\begin{align}\label{eq::SLOPE-ma}
\hat \mu_{SLOPE-ma} = X\left[\sum_{k=1}^K w^{SLOPE}_k \hat\beta_{SLOPE}(q_k)\right],
\end{align}
where the weight vector $w^{SLOPE}=(w^{SLOPE}_1,\cdots,w^{SLOPE}_K)^T\in \mathcal{H}_{K}$ minimizes
\begin{align}
   \|Y - \hat \mu_{SLOPE-ma}\|_2^2 + 2\hat\sigma_{SLOPE}^2 \sum_{k=1}^K w_ks_k,
\end{align}
where $s_k=\|\hat\beta_{SLOPE}(q_k)\|_0$ is the number of active predictors corresponding to solution $\beta_{SLOPE}(q_k)$ and $\hat\sigma_{SLOPE}^2$ is the mean squared residuals corresponding to $\hat\beta_{SLOPE}(q_{cv})$.

Lastly, similar to the lasso case, we define the infeasible SLOPE-optimal solution as 
\begin{align}\label{eq::SLOPE-optimal}
    \hat\mu_{SLOPE-optimal} = X\hat\beta_{SLOPE}(q_{k^*}),
\end{align}
where
\begin{align}
    k^* = \arg\min_k \| X (\hat\beta_{SLOPE}(q_k) - \beta)\|_2^2.
\end{align}
\begin{figure*}[t]
\begin{subfigure}{0.48\textwidth}
\includegraphics[scale=0.45]{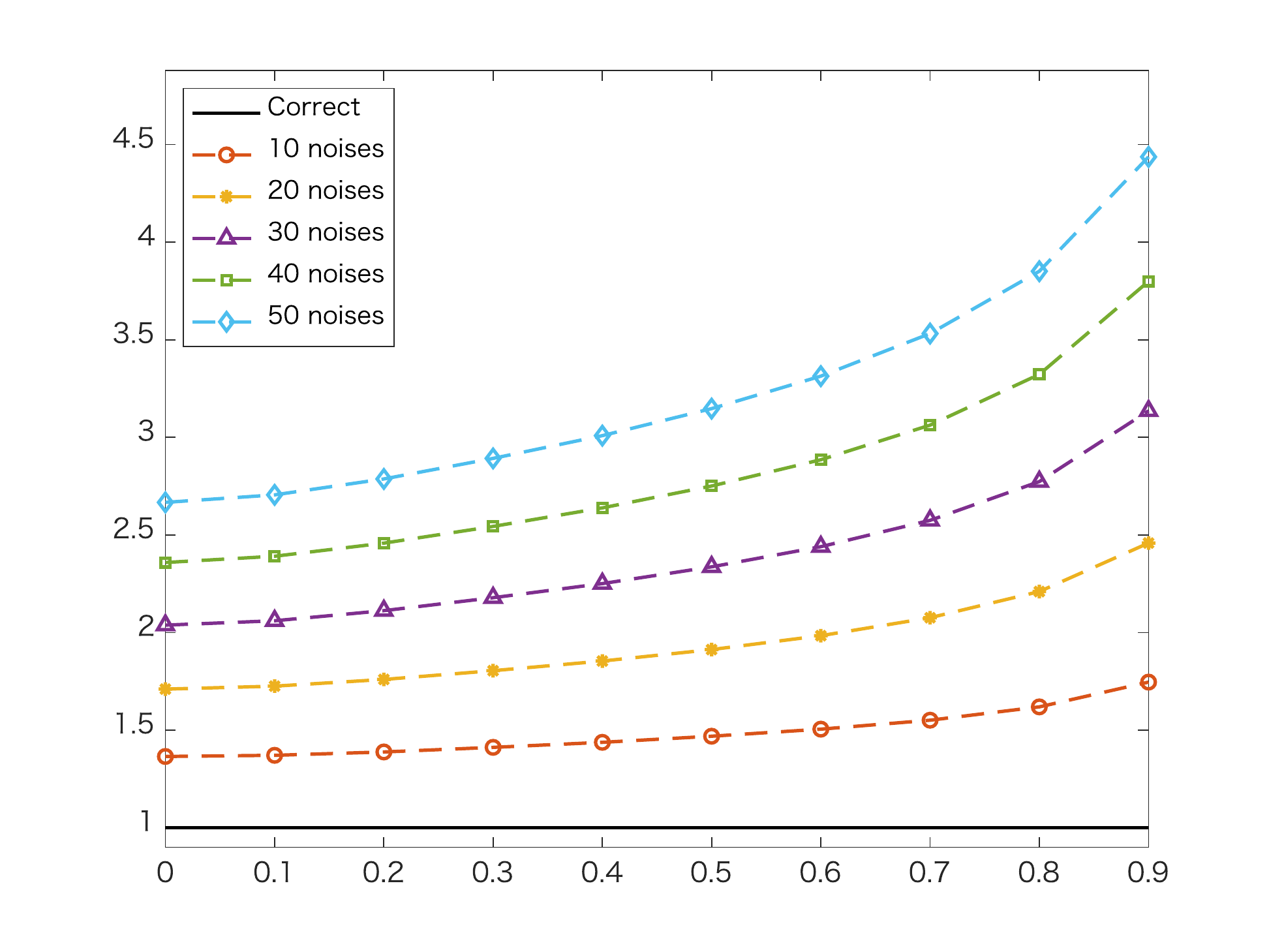}
\vspace{-0.8cm}
    \caption{Relative empirical risk vs. $\rho$ }
\end{subfigure}
\begin{subfigure}{0.48\textwidth}
\includegraphics[scale=0.45]{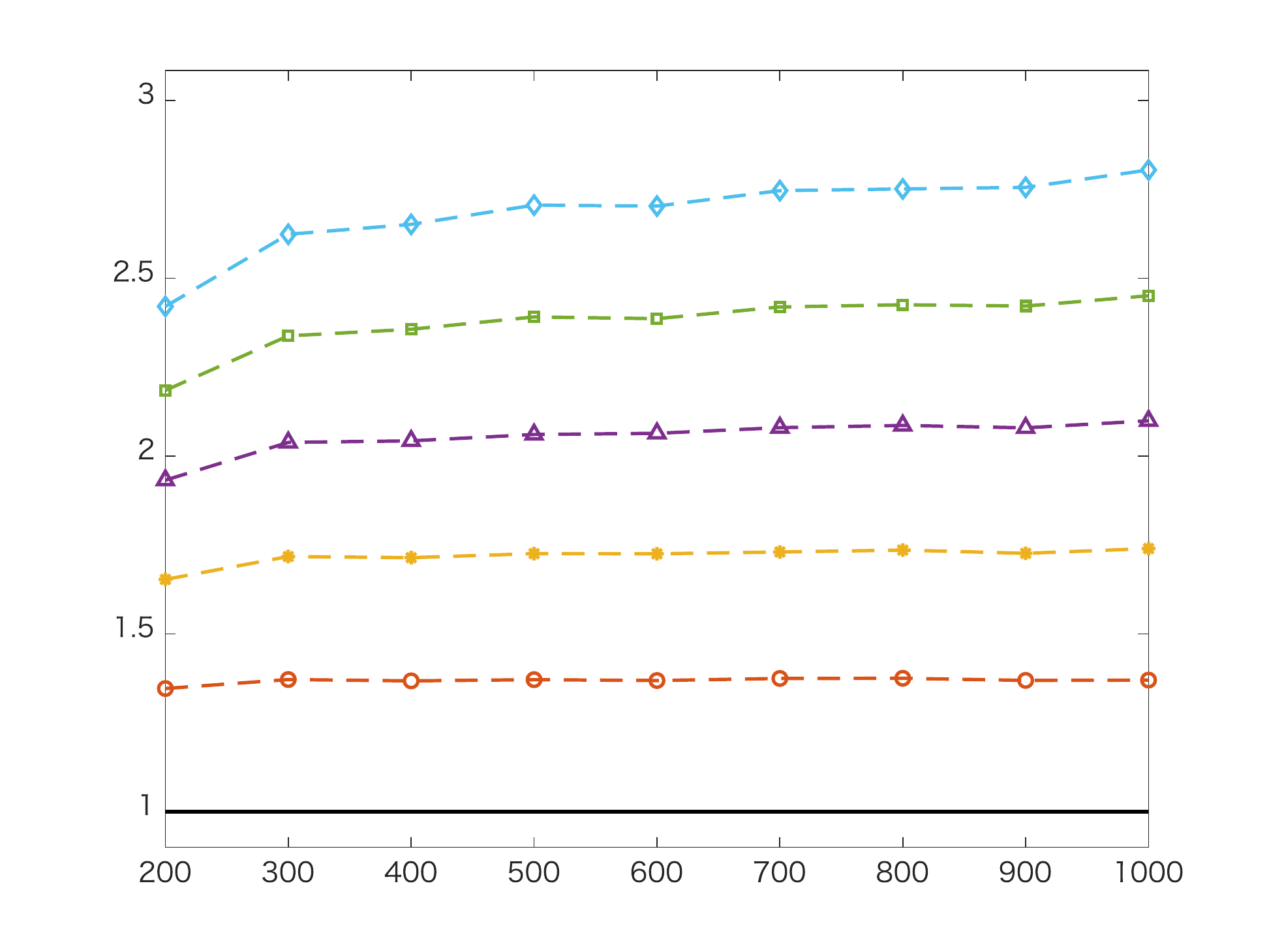}
\vspace{-0.8cm}
    \caption{Relative empirical risk vs. $n$}
\end{subfigure}

\begin{subfigure}{0.48\textwidth}
\includegraphics[scale=0.45]{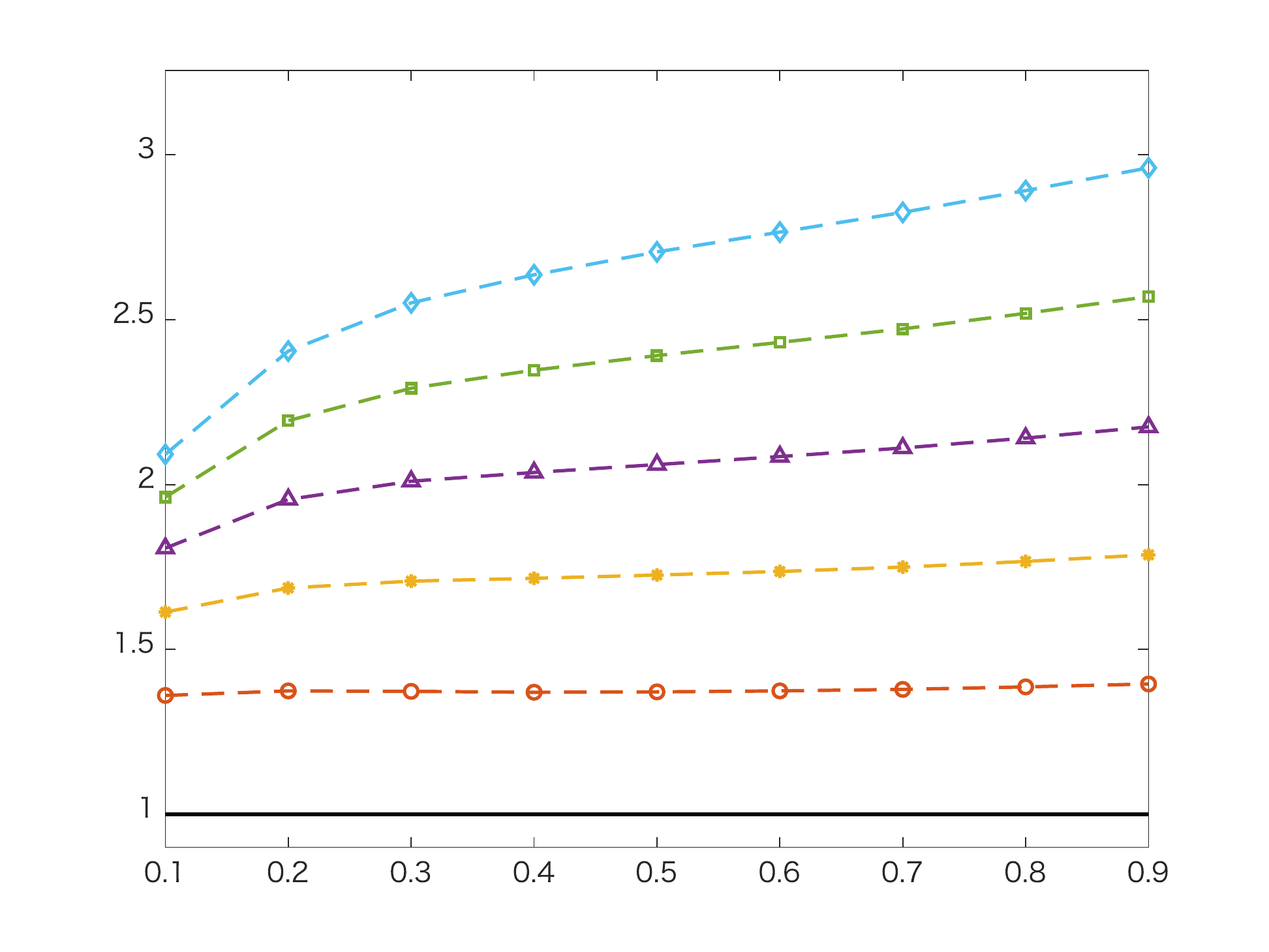}
\vspace{-0.8cm}
    \caption{Relative empirical risk vs. $R^2$ }
\end{subfigure}     
\begin{subfigure}{0.48\textwidth}
\includegraphics[scale=0.45]{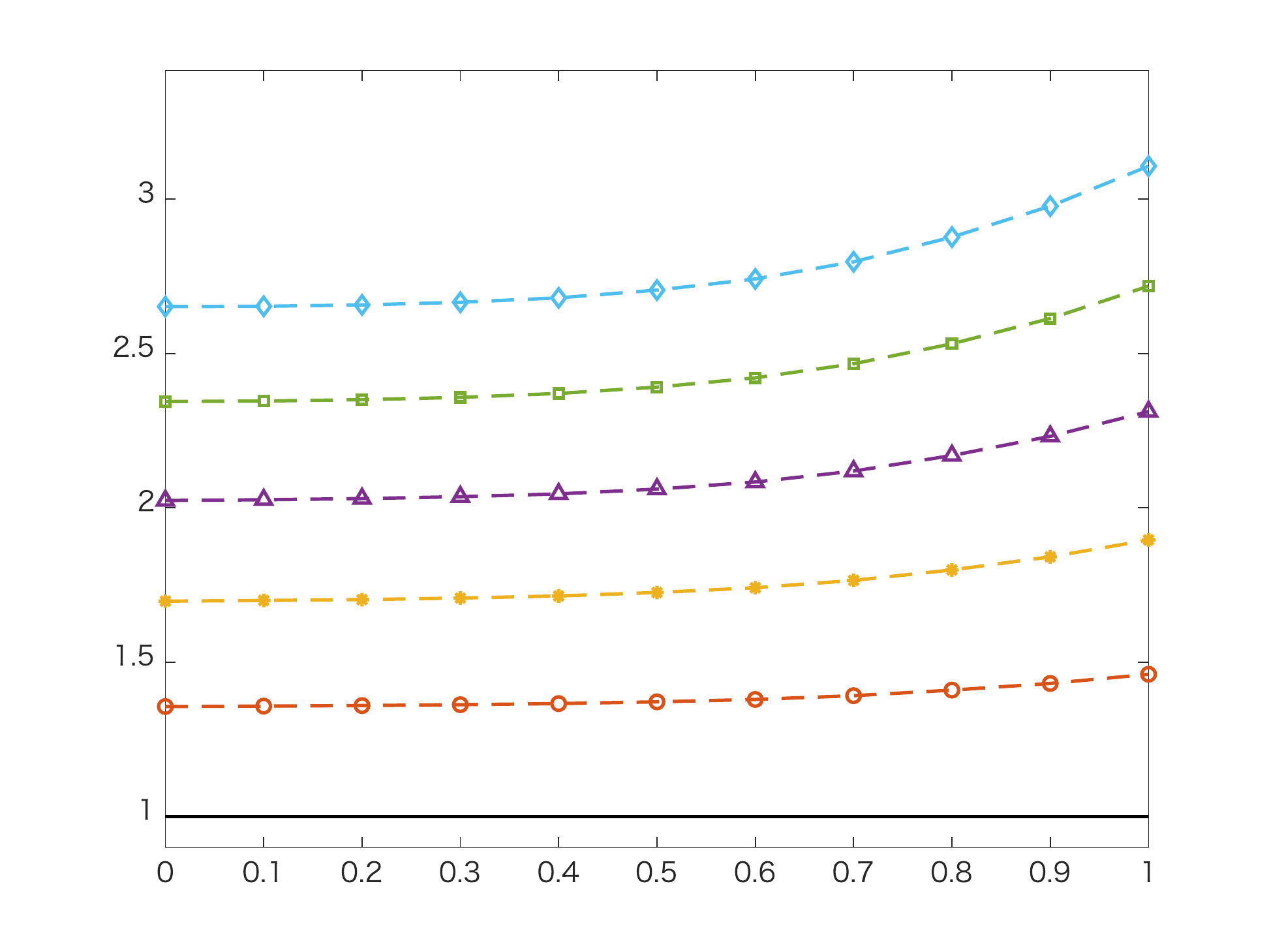}
\vspace{-0.8cm}
    \caption{Relative empirical risk vs. $\delta$ }
\end{subfigure}

\caption{Relative empirical risk over 1000 repetitions when one factor varies for different predictor ordering in MMA. \label{fig::mma-different-ordering}}
\end{figure*}

\section{Simulation\label{sec::simulation}}
In this section, we first conduct a systematic simulation study on the impact of predictor ordering on the performance of nested model averaging (Section \ref{subsec::impact-ordering}), then we compare the proposed nested model averaging methods (lasso-ma and SLOPE-ma) with competing model selection methods (Section \ref{subsec:simu-ma-vs-ms}). 

\subsection{The Impact of Predictor Ordering on Nested Model Averaging}\label{subsec::impact-ordering}
We follow the data generating process as in
 (\ref{eq:lm}). The true parameter $\beta$ is sparse, with only the first 20 entries nonzero. In particular, we set $\beta_j=0$ for $j>20$ and $\beta_{j}=c\cdot j^{-\delta}$ for $j=1,\cdots,20$, where $\delta$ is the coefficient decay rate parameter and $c$ is a constant
to ensure the target population $R^{2}$ value. 
We generate $x=\left[x_{20}',x_{p-20}'\right]'$, where the 20-dimensional vector $x_{20}$ and $(p-20)$-dimensional  vector 
$x_{p-20}$ are independent of each other. Here, $x_{20}\sim N(0,\Sigma)$ with   $\Sigma_{ij}=\rho+(1-\rho)1\{i=j\}$, and 
 $x_{p-20}\sim N(0,I_{p-20})$ with $I_{p-20}$ being an $(p-20)$-dimensional
identity matrix. This represents a compound symmetry (CS) covariance structure among the signals. 

Intuitively, the correct ordering of the predictors for model averaging would have the first 20 predictors (all the important ones) appearing before all the noise variables. To be more specific, we use ``correct ordering" to represent the specific ordering  $(x_1,x_2,\cdots, x_{p})$. Next, we evaluate the case where the order is ``incorrect". In particular, we now put the first $l$ noise predictors before the 20 important predictors while keeping the order unchanged for the remaining noise variables. For example, when $l=10$, the predictor ordering would be $(x_{21},x_{22},\cdots,x_{30},x_1,x_2,\cdots,x_{20},x_{31},x_{32},\cdots,x_p)$. Here, in addition 
to the correct ordering ($l=0$), we consider the incorrect orderings corresponding to $l\in\{10, 20, 30, 40, 50\}$, respectively. 

For each choice of $l$, we apply the MMA method \citep{hansen2007least} with the specified ordering and use the mean squared residuals for the full model with all predictors as $\hat\sigma^2$. The resulting coefficient estimate is denoted by $\hat\beta^{(l)}$, for $l=0$, 10, 20, 30, 40, and 50. We then calculate the \emph{in-sample loss} $L_n(\hat\beta^{(l)}) = \| X(\hat{\beta}^{(l)}-\beta)\|_2^{2}$. Each experiment is repeated for 1000 times and the average is interpreted as the \emph{empirical risk} $\hat R_n^{(l)}$. Lastly, we calculate the \emph{relative empirical risk} of each case relative to $l=0$ as $\hat R_n^{(l)}/\hat R_n^{(0)}$. 

Regarding the parameter setting, we identify four factors that could affect the relative empirical risk.
\begin{enumerate}[(a)]
    \item Common correlation coefficient $\rho$ among important predictors.
    \item Sample size $n$.
    \item The target population $R^2$ value. 
    \item The coefficient decay rate $\delta$ for the regression coefficients. 
\end{enumerate}
Then, we consider in a total of four experiments. In each experiment, one of
the four factors $\rho$, $n$, $R^2$, and $\delta$ is changing, while the
remaining factors are fixed according to the following values:  $\rho=0.1$, $n=500$, $R^{2}=0.5$,  $\delta=0.5$. Here, the total number of predictors is fixed at $p=150$. The range of the factors is specified as follows.
\begin{enumerate}[(a)]
\item $\rho\in\{0:0.1:0.9\}$. Note that throughout this paper, we use $0:0.1:0.9$ to represent the evenly spaced sequence
from 0 to 0.9 with increment 0.1, i.e., $\{$0, 0.1, 0.2, 0.3, 0.4, 0.5, 0.6, 0.7, 0.8, 0.9$\}$.
\item $n\in\{200:100:1000\}$. 
\item $R^{2}\in\{0.1:0.1:0.9\}$. 
\item $\delta\in\{0:0.1:1\}$. 
\end{enumerate}
The results are summarized in Figure \ref{fig::mma-different-ordering}, panels (a)-(d),
respectively. It is clear that across all four experiments, putting noise predictors before the important predictors always leads to a larger empirical risk, and the relative empirical risk gets larger as the number of such noise predictors increases. For example, when we have $50$ noise variables ordered before the important variables, we can see a three to four-fold increase in the empirical risk, which is quite substantial.  This simulation indicates that it is vital that we have a high-quality ordering in the nested model averaging in order to reduce the empirical risk. Next, we show that the ordering based on the solution paths generated by lasso and SLOPE could lead to great performance under various settings. 

\subsection{Comparing nested model averaging methods with model selection\label{subsec:simu-ma-vs-ms}}
In this section, we conduct extensive simulations to evaluate the performance of two model averaging estimators, namely lasso-ma and SLOPE-ma, by comparing them with lasso-ols, SLOPE-ols, lasso-optimal, and SLOPE-optimal. We want to reiterate that lasso-optimal and SLOPE-optimal are \emph{oracle-type} estimators that are not feasible in practice. By considering the infeasible lasso-optimal and SLOPE-optimal estimates, it is not necessary to consider different tuning parameter selection methods for choosing the \emph{best} performing solution on the path. 

We follow the data generation process as in \eqref{eq:lm}. We assume the true parameter $\beta$ is sparse with only the first $s$ entries nonzero. In particular, we set $\beta_j=0$ for $j>s$ and $\beta_j = c\times j^{-\delta}$ for $j=1\cdots, s$, where $\delta$ is the coefficient decay rate and $c$ is a constant to ensure the target population $R^2$ value. We generate $x=\left[x_{s}',x_{p-s}'\right]'$, where the $s$-dimensional vector $x_{s}$ represents the signals and $(p-s)$-dimensional vector represents
$x_{p-s}$ the noises. We assume $x\sim N(0,\Sigma)$  with the following two commonly used correlation structures among the $p$ predictors, including both signals and noises. 
\begin{itemize}
    \item (Compound Symmetry). We assume the correlation among any two predictors is always $\rho$, i.e., $\Sigma_{ij}=\rho+(1-\rho)1\{i=j\}$, for $i,j=1,\cdots,p$.
    \item (Auto Regressive). We assume the correlation among two predictors decays exponentially as a function of the differences between their indices in absolute value, i.e., $\Sigma_{ij}=\rho^{|i-j|}$, for $i,j = 1,\cdots, p$.
\end{itemize}
In both correlation structures, the parameter $\rho$ controls the strength of correlation. When we fix the value of $\rho$, the compound symmetry correlation structure assumes an overall stronger correlation than the auto regressive structure. We will investigate both correlation structures. 

For each correlation structure, we study the impact of sample size $n$, population $R^2$,  total number of predictors $p$,  correlation  parameter $\rho$, the number of important predictors $s$ and the coefficient decay rate $\delta$. In particular, we 
consider the following six experiments where we vary one factor while keeping the remaining factors fixed according to the following values: $n=500$, $R^2=0.5$,  $p=600$, $\rho=0.1$, $s=100$,  and $\delta=0.5$. The range of the factors are specified as follows.
\begin{enumerate}[(a)]
 \item Sample size $n\in\{200:100:1000\}$. 
    \item The population $R^2\in \{0.1:0.1:0.9\}$. 
   
    \item Total number of predictors $p\in\{200:100:1000\}$.
    \item Correlation parameter $\rho \in \{0:0.1:0.9\}$. 
    \item Important predictors number $s\in \{50:50:500\}$. 
    \item The coefficient decay rate $\delta \in\{0:0.1:1\}$. 
\end{enumerate}
For each experiment and every method, we first calculate the estimate $\hat\beta$ and its corresponding  \emph{in-sample loss} $L_n(\hat\beta) = \| X(\hat{\beta}-\beta)\|_2^{2}$. Each experiment is repeated for 1000 times and the average is interpreted as the \emph{empirical risk}: $\hat R_n^{lasso-ma}$, $\hat R_n^{SLOPE-ma}$, $\hat R_n^{lasso-ols}$, $\hat R_n^{SLOPE-ols}$, $\hat R_n^{lasso-optimal}$ and $\hat R_n^{SLOPE-optimal}$. 


\begin{table*}[ht]
\begin{threeparttable}
\caption{The average mean squared prediction error\label{tb::crime}}
\begin{tabular}{lllllll}
\hline
$tr_p$&lasso-cv&lass-ols&lasso-ma&SLOPE-cv&SLOPE-ols&SLOPE-ma\\ \hline
0.3&163.76(0.2) &169.45(0.29) &162.51(0.2) &172.37(0.19) &166.68(0.23) &172.48(0.2)\\
0.4&160.58(0.19) &164.49(0.25) &158.8(0.19) &168.82(0.19) &162.42(0.21) &168.63(0.19)\\
0.5&159.03(0.23) &162.24(0.27) &156.85(0.23) &166.4(0.23) &159.95(0.25) &166.27(0.23)\\
0.6&157.91(0.29) &160.48(0.31) &155.55(0.29) &164.57(0.28) &158.51(0.3) &164.33(0.28)\\
0.7&156.13(0.35) &157.96(0.36) &154.02(0.34) &162.88(0.34) &157.2(0.36) &162.66(0.34)\\
0.8&156.03(0.47) &157.17(0.48) &154.48(0.46) &163.2(0.44) &157.93(0.47) &163.04(0.45)\\
0.9&153.38(0.65) &154.22(0.66) &152.4(0.64) &160.55(0.61) &156.02(0.65) &160.4(0.62)\\ \hline
\end{tabular}
\begin{tablenotes}
\item The number of repetitions is $500$ for various training proportions. The standard errors are in \item parentheses.
    \end{tablenotes}
\end{threeparttable}
\end{table*}

\begin{center}

\begin{figure*}
\begin{subfigure}{0.48\textwidth}
\includegraphics[scale=0.48]{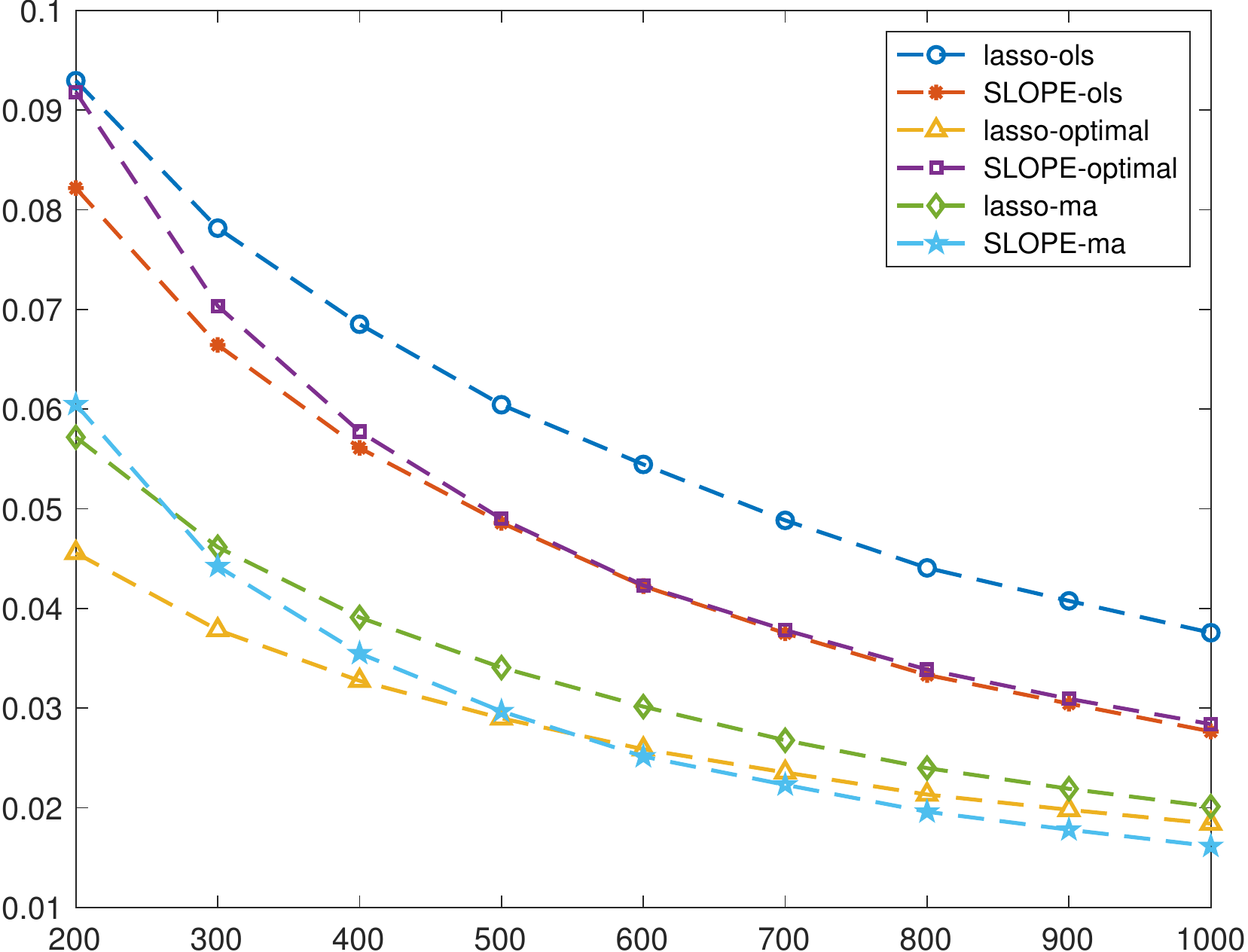}
    \caption{Empirical risk vs. $n$}
    \vspace{0.3cm}
\end{subfigure}
\begin{subfigure}{0.48\textwidth}
\includegraphics[scale=0.48]{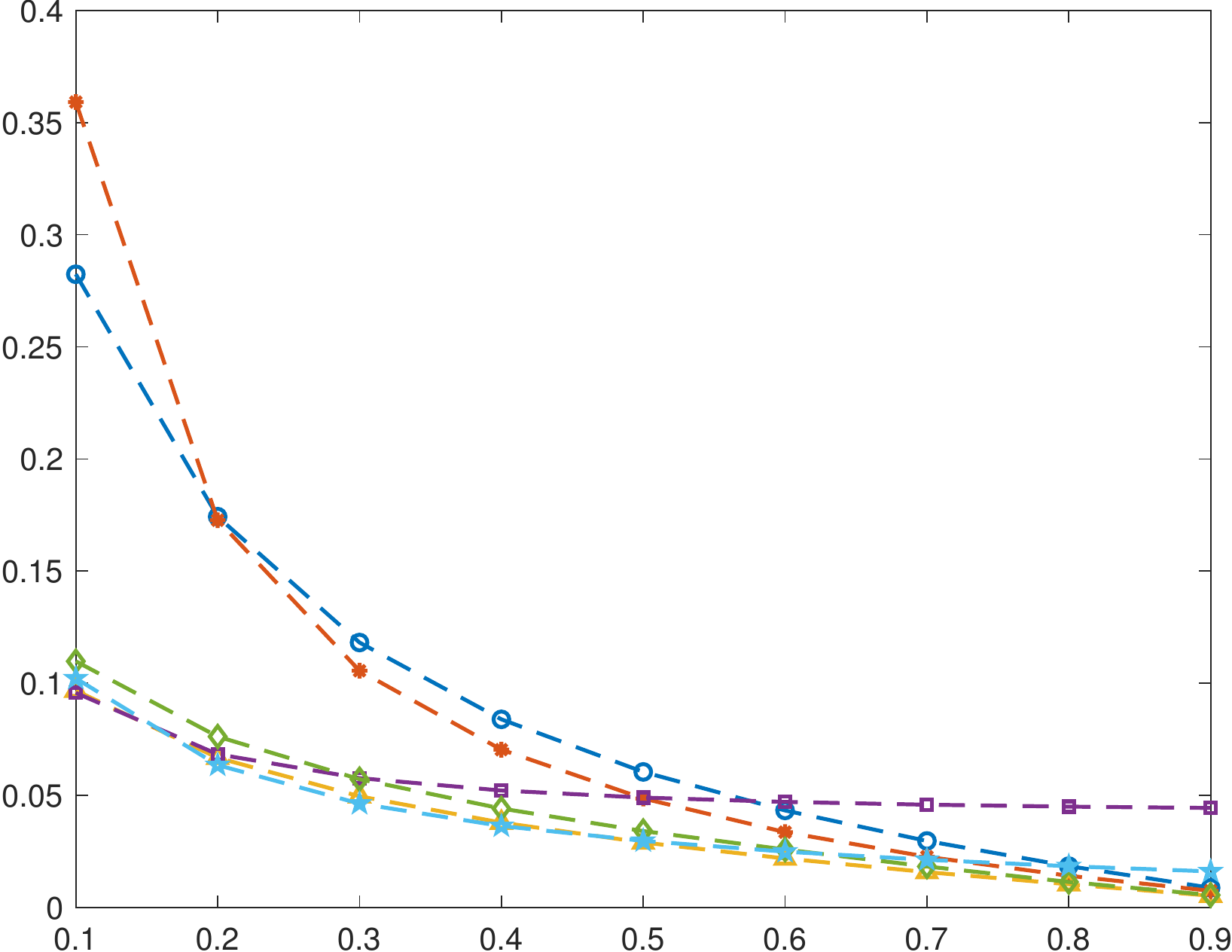}
    \caption{Empirical risk vs. $R^2$ }
    \vspace{0.3cm}
\end{subfigure}
\begin{subfigure}{0.48\textwidth}
\includegraphics[scale=0.48]{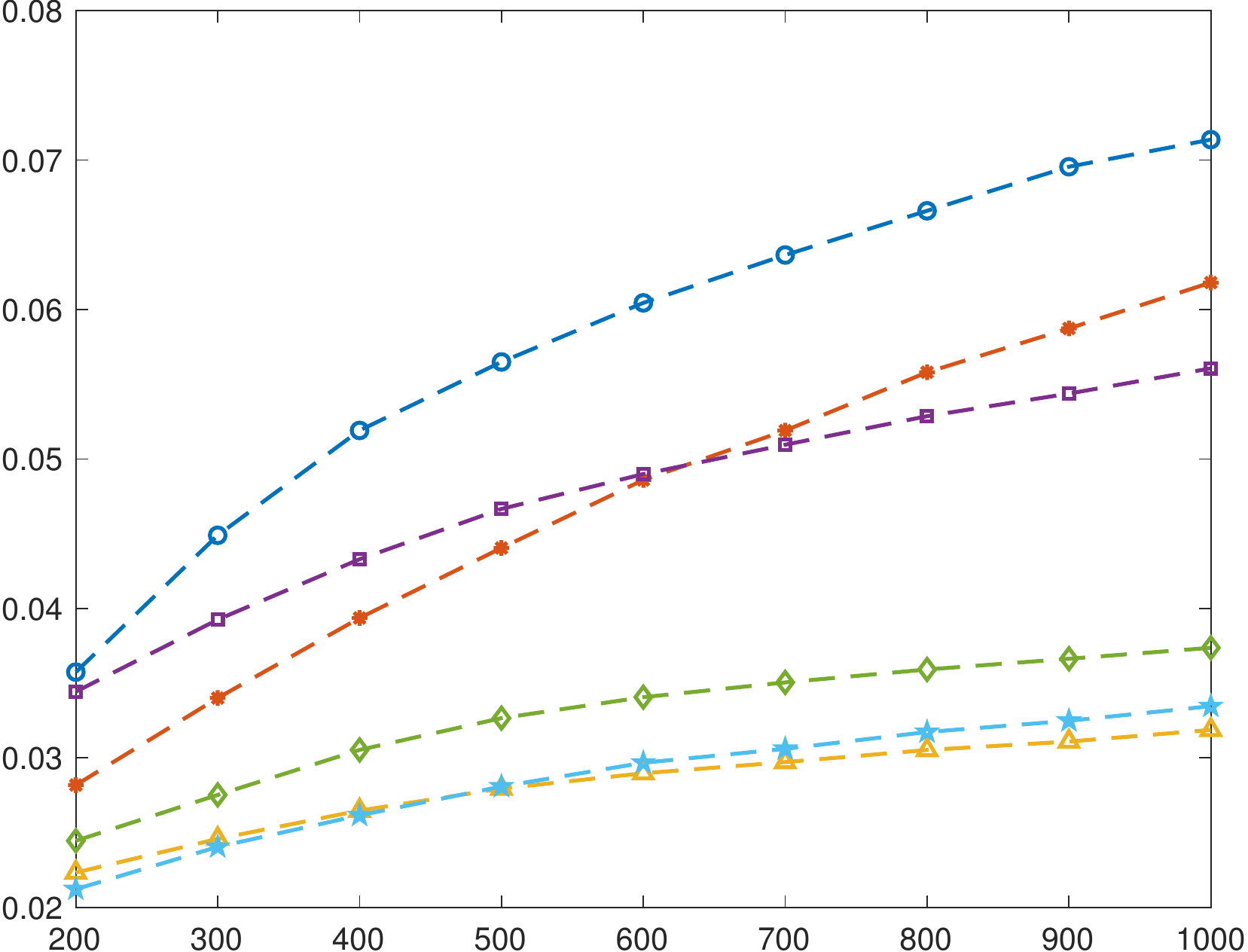}
    \caption{Empirical risk vs. $p$ }
    \vspace{0.3cm}
\end{subfigure}
\begin{subfigure}{0.48\textwidth}
\includegraphics[scale=0.48]{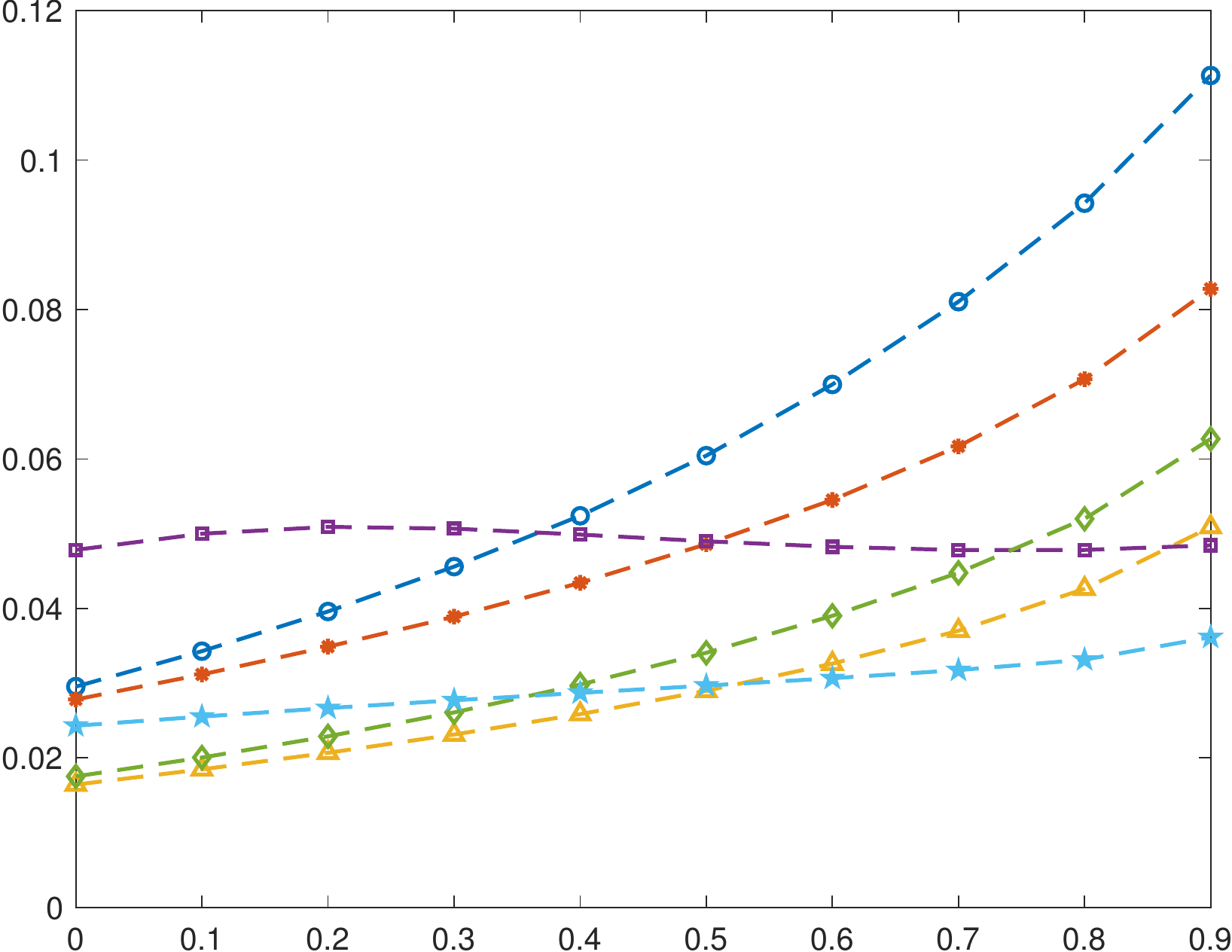}
    \caption{Empirical risk vs. $\rho$ }\vspace{0.3cm}
\end{subfigure}

\begin{subfigure}{0.48\textwidth}
\includegraphics[scale=0.48]{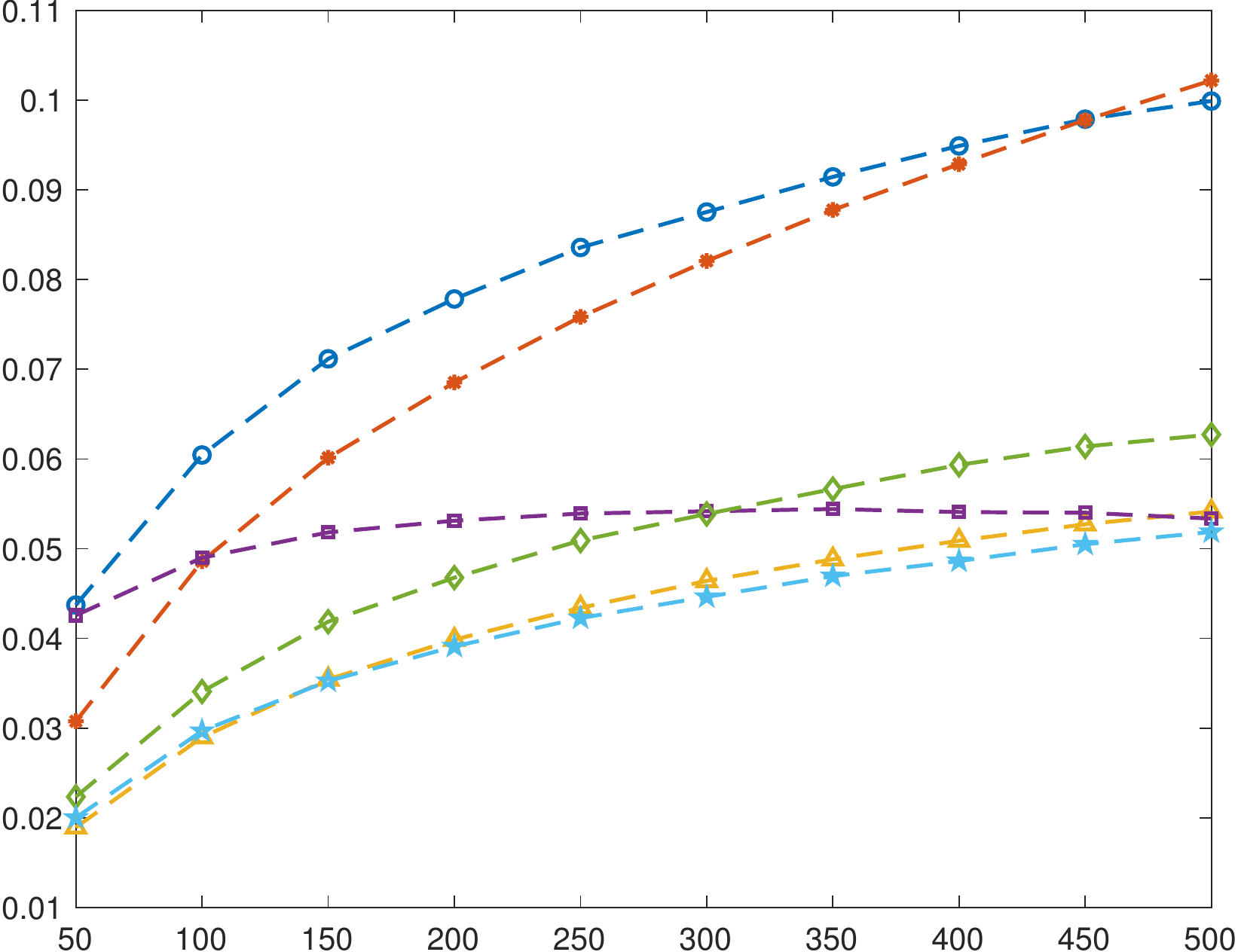}
    \caption{Empirical risk vs. $s$ }
\end{subfigure}      
    \begin{subfigure}{0.48\textwidth}
\includegraphics[scale=0.48]{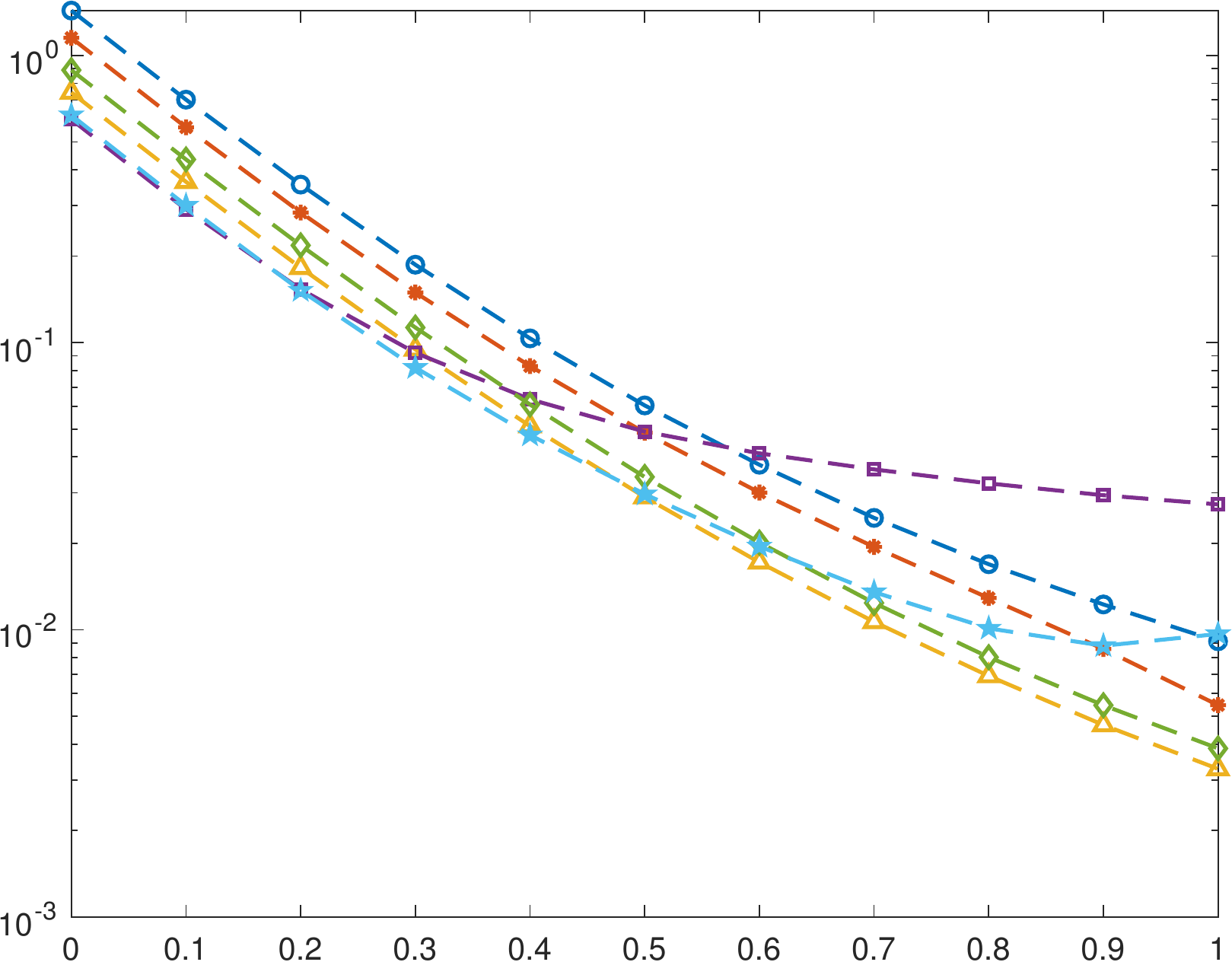}
    \caption{Empirical risk vs. $\delta$ }
\end{subfigure}

\caption{Empirical risk over 1000 repetitions  under the Auto Regressive correlation structure. \label{fig:simulation-AR}}

\end{figure*}
\end{center}

\begin{center}

\begin{figure*}[t]
\begin{subfigure}{0.48\textwidth}
\includegraphics[scale=0.48]{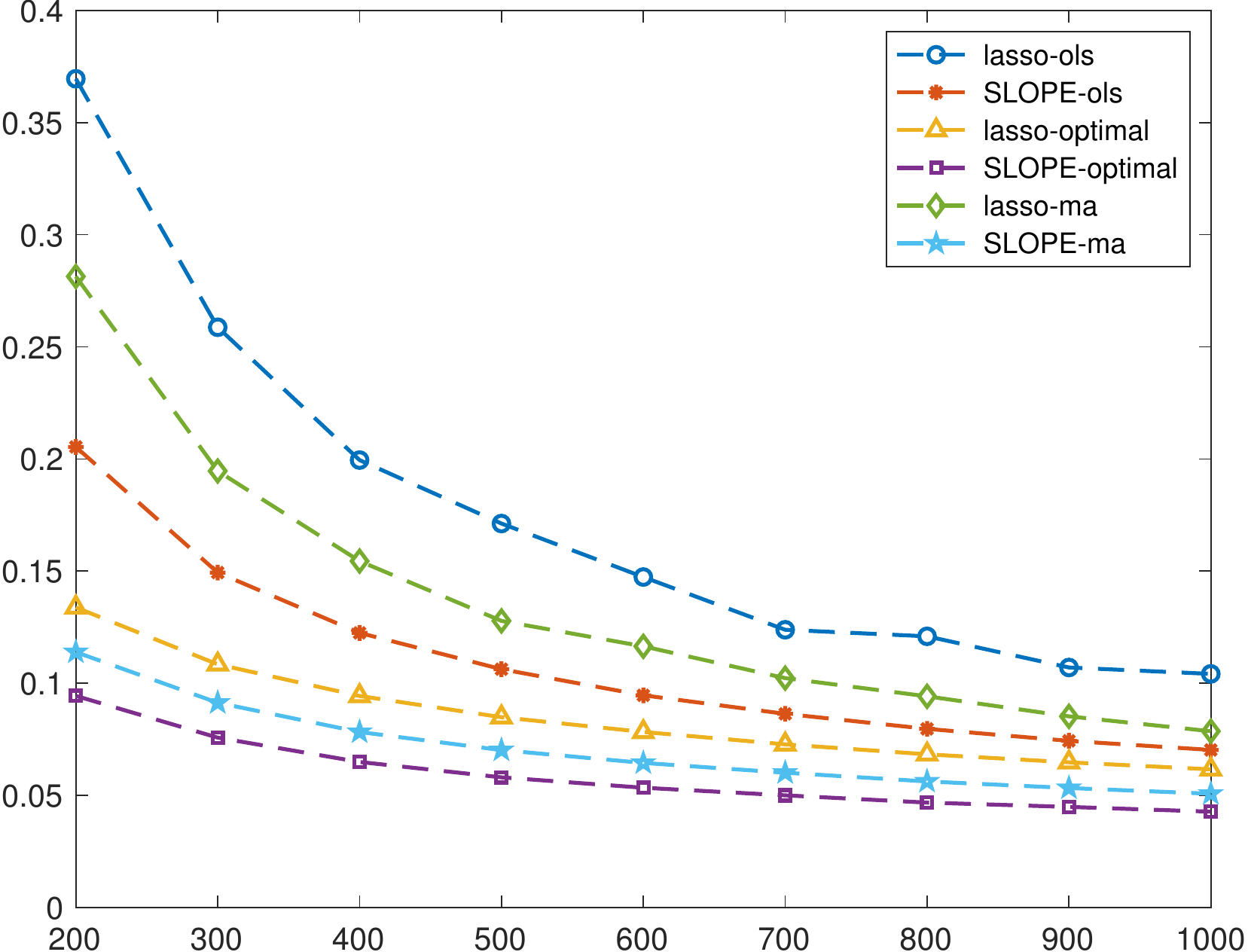}
    \caption{Empirical risk vs. $n$}\vspace{0.3cm}
\end{subfigure}
\begin{subfigure}{0.48\textwidth}
\includegraphics[scale=0.48]{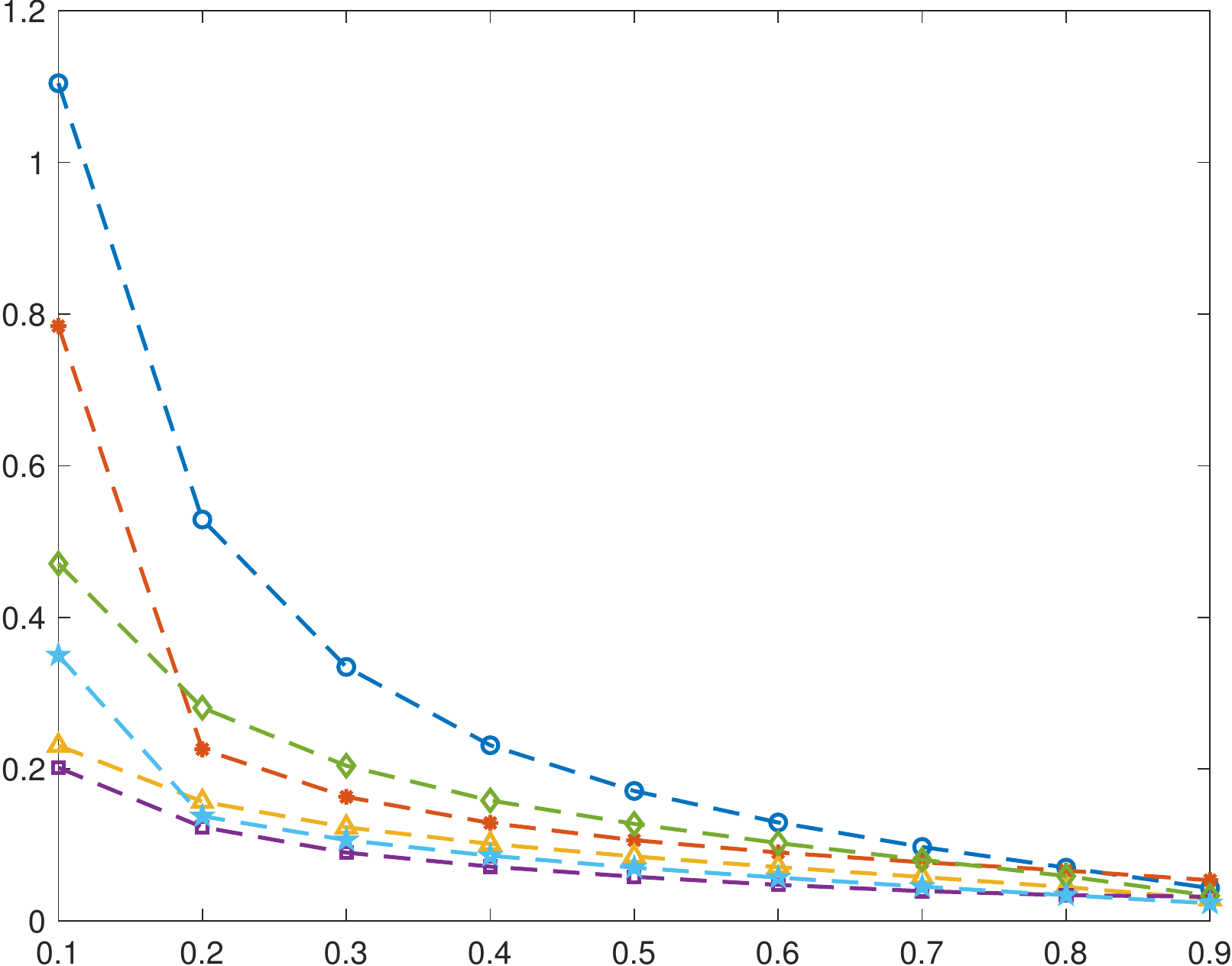}
    \caption{Empirical risk vs. $R^2$ }\vspace{0.3cm}
\end{subfigure}
\begin{subfigure}{0.48\textwidth}
\includegraphics[scale=0.48]{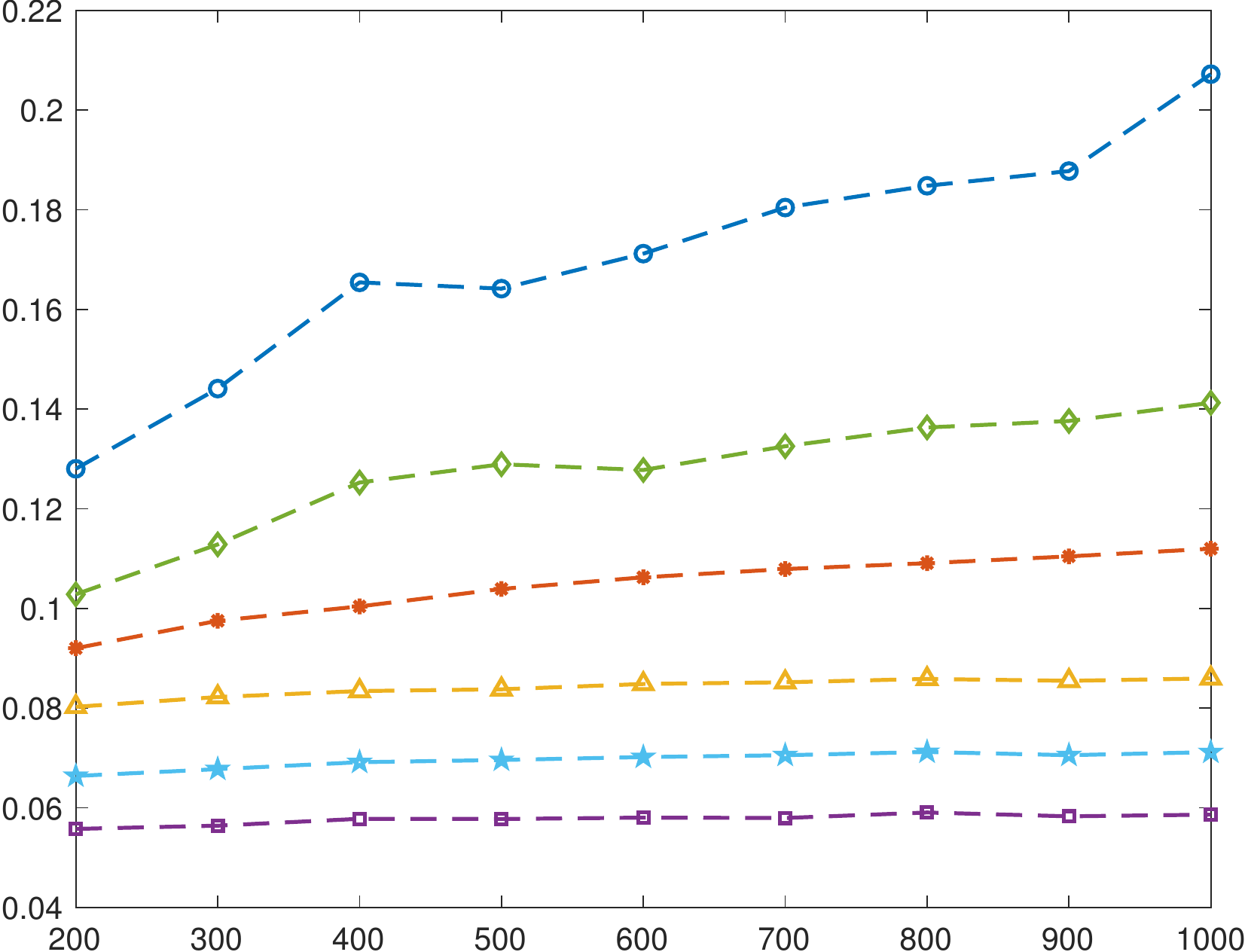}
    \caption{Empirical risk vs. $p$ }\vspace{0.3cm}
\end{subfigure}
\begin{subfigure}{0.48\textwidth}
\includegraphics[scale=0.48]{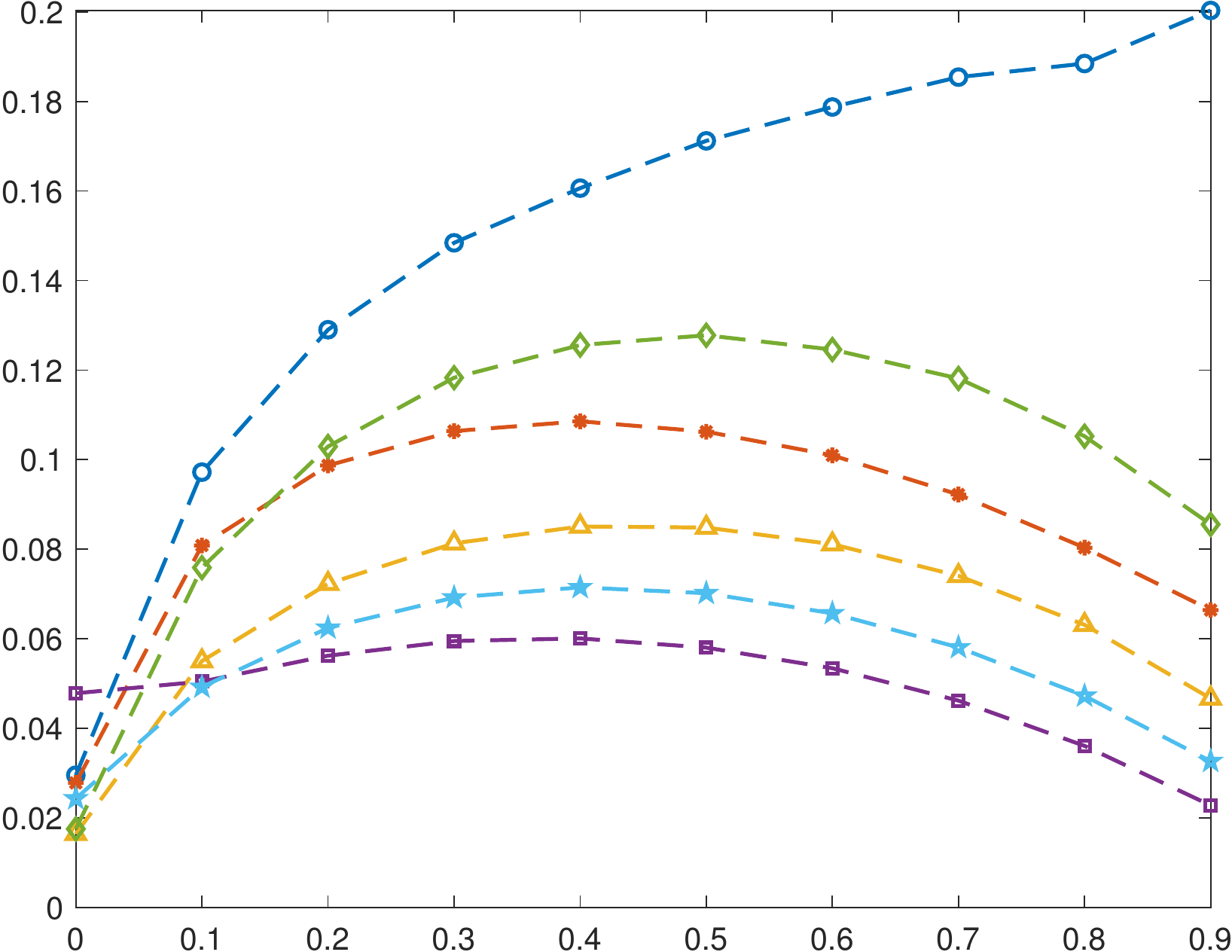}
    \caption{Empirical risk vs. $\rho$ }\vspace{0.3cm}
\end{subfigure}

\begin{subfigure}{0.48\textwidth}
\includegraphics[scale=0.48]{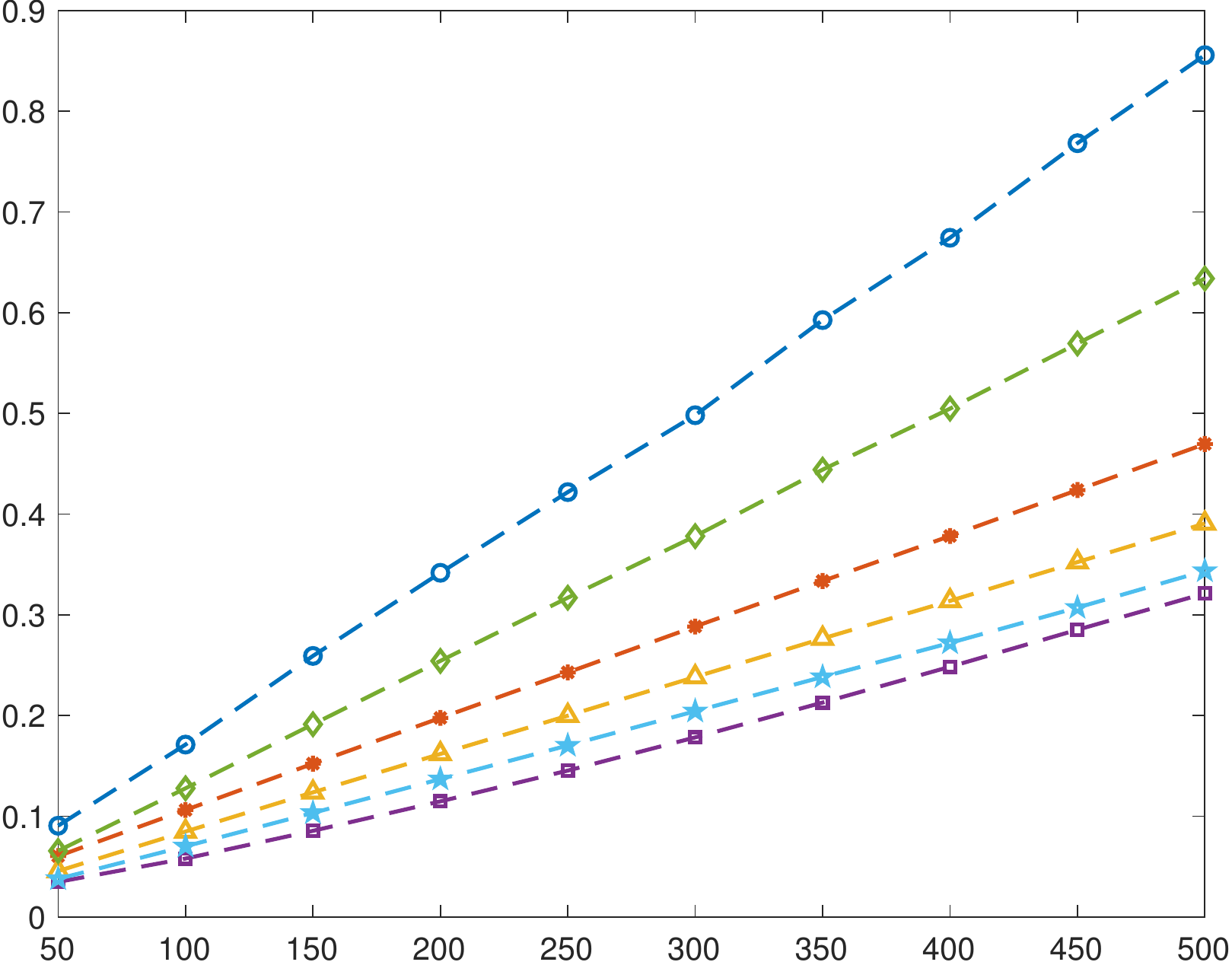}
    \caption{Empirical risk vs. $s$ }
\end{subfigure}      
    \begin{subfigure}{0.48\textwidth}
\includegraphics[scale=0.48]{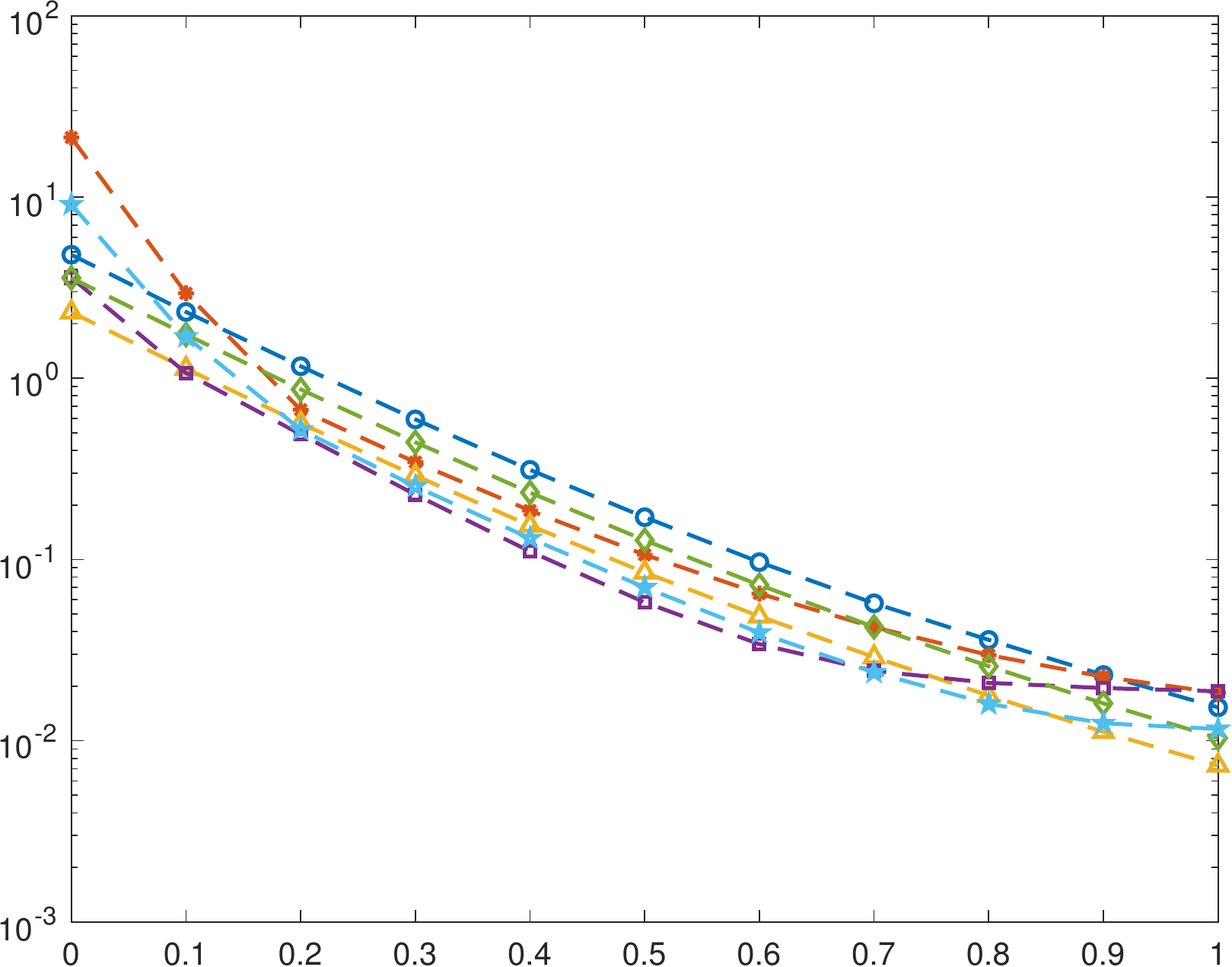}
    \caption{Empirical risk vs. $\delta$ }
\end{subfigure}

\caption{Empirical risk over 1000 repetitions  under the Compound Symmetry correlation structure. \label{fig:simulation-CS}}
\end{figure*}
\end{center}

The results of the six experiments are summarized in Figures \ref{fig:simulation-AR} and \ref{fig:simulation-CS} for the Auto Regressive and Compound Symmetry correlation settings, respectively. For each setting,  panels (a)-(f) correspond to the corresponding experiment. We have the following observations.

First, we focus on the Auto Regressive correlation setting in Figure \ref{fig:simulation-AR}. The following observations are made.
\begin{itemize}
    \item Across almost all parameter combinations considered, the model averaging based methods (lasso-ma and SLOPE-ma) improve over their model selection counterparts (lasso-ols and SLOPE-ols), respectively. Sometimes, the improvement could be quite substantial. For example, the empirical risk of SLOPE-ma is less than one third of that of SLOPE-ols in panel (b) when $R^2=0.1$. 
\item  It is worth noting that the model averaging methods (lasso-ma and SLOPE-ma) even outperforms the infeasible model selection methods (lasso-optimal and  SLOPE-optimal) for some of the settings. This shows that the process of model averaging improves over any individual estimate uniformly. We would like to provide some intuitions as follows.
 Suppose
$\beta$ is known, we have
\begin{align*}
  \min_{k}\|X(\hat{\beta}_{lasso}(\lambda_{k})-\beta)\|_{2}^{2}
= & \underset{w^{lasso}}{\min}\sum_{k=1}^{K}w_{k}^{lasso}\|X(\hat{\beta}_{lasso}(\lambda_{k})-\beta)\|_{2}^{2}\\
\geq & \underset{w^{lasso}}{\min}\|X(\sum_{k=1}^{K}w_{k}^{lasso}\hat{\beta}_{lasso}(\lambda_{k})-\beta)\|_{2}^{2}.
\end{align*}
 Indeed, the candidate space in which we search the lasso-ma solution
$\hat{\mu}_{lasso-ma}$ is not restricted on but includes all the solutions
of the lasso solution path. Consequently, $\hat{\mu}_{lasso-ma}$
has potential to over-perform the lasso-optimal solution $\hat{\mu}_{lasso-optimal}$.
Obviously, the SLOPE-optimal solution \\ $\hat{\mu}_{SLOPE-optimal}$  is in the same situation. The
theoretical analysis of the comparison between the ma solutions
and the corresponding infeasible optimal solutions is out of our scope. As a relative
result, \citet{Hansen2014Model} showed that the OLS solution of any
single linear models has a larger risk than that of the MMA. 
\item Another interesting finding is that the lasso-ols does not behave well throughout the six experiments, possibly because the number of important predictors is relatively large compared with the dimensionality and the correlation among predictors is substantial, which may violate the irrepresentable condition required for  model selection consistency of lasso \citep{zhao2006model}. 
\item Regarding the two model averaging methods, lasso-ma and SLOPE-ma perform similarly in panel (b). SLOPE-ma  outperforms lasso-ma for panels (a), (c) and (e).  From panel (d),  lasso-ma outperforms SLOPE-ma when the correlation parameter $\rho$ is small, and SLOPE-ma has a smaller empirical risk than that of lasso-ma when $\rho$ is large. From panel (f), lasso-ma can reduce the empirical risk to a smaller level than SLOPE-ma when the coefficient decay rate $\delta$ is large. This is intuitive as lasso tends to work better when the problem is sparser. 
\end{itemize}
Let's now move to the corresponding risk for the Compound Symmetry correlation case presented in Figure \ref{fig:simulation-CS}. To avoid repetition, we will only highlight the new findings under this scenario. 
\begin{itemize}
    \item Generally speaking, lasso-ma has a much worse performance compared with the corresponding setting under the Auto Regressive scenario. One possible reason is that the irrepresentable condition, required for achieving model selection consistency for lasso, is not satisfied under the Compound Symmetry correlation setting. Because of this, the variable ordering from the lasso solution path may not be the correct ordering, therefore impacting the behavior according to Section \ref{subsec::impact-ordering}.
    \item SLOPE-ma along with SLOPE-optimal are the best performing methods across nearly all settings. SLOPE-ma performs slightly better than SLOPE-optimal in panels (d) and (f) when the correlation is 0 and when the decay parameter $\delta=1$, respectively. 
\end{itemize}

This array of simulation experiments show that the model averaging methods tend to improve over the model selection methods. Choosing between lasso-ma and SLOPE-ma would mainly depend on the correlation structure among predictors and the coefficient distributions.

\section{Real Data Analysis\label{sec:real}}

In this section, we investigate the behaviors of the proposed methods via a crime data set \citep{adams1992census, redmond2002data}. The data set combines socioeconomic data from the 1990 US Census, law enforcement data from the 1990 US LEMAS survey, and crime data from the 1995 FBI UCR. It was downloaded from the UCI Machine Learning Repository at \url{http://archive.ics.uci.edu/ml/datasets/communities+and+crime}. The goal is to predict the \emph{Per Capita Violent Crimes} using covariates involving the community, such as the percent of the population considered urban, and the median family income, and involving law enforcement, such as per capita number of police officers, and percent of officers assigned to drug units. The per capita violent crimes variable was calculated using population and the sum of crime variables considered violent crimes in the United States: murder, rape, robbery, and assault. 

We first remove the covariates that have missing values, leaving us with $p = 99$ covariates and $n=1,994$ observations in total. Next, we randomly split the data into a training set ($n_{tr}$ observations) and test sets ($n_{te}$ observations) with their sizes to be specified. Let lasso-cv and slope-cv denote the original lasso and slope methods whose tuning parameter $\lambda$ is selected by 10-fold cross-validation. For each of the six methods (lasso-cv, lasso-ols, lasso-ma, slope-cv, slope-ols, slope-ma), we obtain the estimate $\hat{\beta}$ of the regression coefficients using the training set and calculate the following mean squared prediction error on the test set.
\begin{align}
    MSPE_{\hat\beta} = \frac{1}{n_{te}}\sum_{i=1}^{n_{te}}(Y^{(te)}_i-\hat\beta^T X^{(te)}_i)^2.
\end{align}

We vary the training proportion $tr$ from 0.3 to 0.9 with increment 0.1. Then, we set $n_{tr} = \lceil n\times tr\rceil$ and $n_{te}=n-n_{tr}$. The average and standard error among 500 repetitions are reported in Table \ref{tb::crime}.

From Table \ref{tb::crime}, it is clear that under most training proportions, the model averaging based method lasso-ma leads to the smallest MSPE across all training proportions. It is worth noting that lasso-ols has an inflated MSPE compared with lasso-cv, while SLOPE-ols's MSPE is smaller than that of SLOPE-cv. One possible explanation is that lasso-cv tends to select more predictors than SLOPE-cv, resulting in the ols setup unstable for lasso-ols. 
\section{Discussion\label{sec:discussion}}
In this paper, we provide two nested model averaging methods, namely lasso-ma and SLOPE-ma, for high-dimensional linear regression.  By taking advantage of the high-quality ordering of lasso and SLOPE solution paths, we show the nested model averaging estimators  works well across a wide range of simulated scenarios and a real data set. We also conducted a detailed simulation study to quantify the impact of predictor ordering on the performance of nested model averaging. 

There are a few open questions. First, it would be interesting to study the impact of predictor ordering from a theoretical perspective. In particular, we would like to quantify how the in-sample loss and the risk change as a function of the incorrect degree of the predictor ordering. Second, the proposed methodology can be easily extended to more general models, including the generalized linear models and Cox model.  Another appealing direction is to develop the theoretical properties of lasso-ma and SLOPE-ma. 
Lastly, model averaging with other types of solution path algorithms (e.g., step-wise regression, forward regression) are worth further investigation in the regime of high-dimensional regression. 

\bibliographystyle{ACM-Reference-Format}
\bibliography{MASP}

\end{document}